\documentclass[aps,prl,reprint,superscriptaddress,longbibliography]{revtex4-1}
\usepackage{blindtext}
\pdfoutput=1

\usepackage{epstopdf}
\usepackage{graphicx}
\usepackage{titlesec}
\newcommand{\tab}{\hspace*{2em}}
\newcommand{\mytilde}{\raise.17ex\hbox{$\scriptstyle\mathtt{\sim}$}}
\newcommand{\RNum}[1]{\uppercase\expandafter{\romannumeral #1\relax}}
\usepackage[margin=.75in]{geometry}

\begin{document}
\title{{\bf Cavity-enhanced measurements of defect spins in silicon carbide}}
\author{Greg Calusine}
\affiliation{Department of Physics, University of California, Santa Barbara, California 93106, USA}
\author{Alberto Politi}
\affiliation{Department of Physics, University of California, Santa Barbara, California 93106, USA}
\affiliation{School of Physics and Astronomy, University of Southampton, Southampton SO17 1BJ, United Kingdom.}
\author{David D. Awschalom}
\email{awsch@uchicago.edu}
\affiliation{Department of Physics, University of California, Santa Barbara, California 93106, USA}
\affiliation{Institute for Molecular Engineering, University of Chicago, Chicago, Illinois 60637, USA}

\begin{abstract} 
{\tab}{\bf The identification of new solid-state defect qubit candidates in widely used semiconductors has the potential to enable the use of nanofabricated devices for enhanced qubit measurement and control operations.  In particular, the recent discovery of optically active spin states in silicon carbide thin films offers a scalable route for incorporating defect qubits into on-chip photonic devices.  Here we demonstrate the use of 3C silicon carbide photonic crystal cavities for enhanced excitation of color center defect spin ensembles in order to increase measured photoluminescence signal count rates, optically detected magnetic resonance signal intensities, and optical spin initialization rates.  We observe up to a factor of 30 increase in the photoluminescence and ODMR signals from Ky5 color centers excited by cavity resonant excitation and increase the rate of ground-state spin initialization by approximately a factor of two.  Furthermore, we show that the small excitation mode volume and enhanced excitation and collection efficiencies provided by the structures can be used to study inhomogeneous broadening in defect qubit ensembles.  These results highlight some of the benefits that nanofabricated devices offer for engineering the local photonic environment of color center defect qubits to enable applications in quantum information and sensing.}
\end{abstract}

\maketitle

\section{\large \RNum{1}. INTRODUCTION}
{\tab}Electronic spins associated with color center defects in silicon carbide (SiC) show promise as a potential component for solid-state quantum technologies due to their combination of long coherence times,\cite{Christle2014} room-temperature operation,\cite{Koehl2011} and a host material for which mature growth\cite{Powell2006} and fabrication protocols exist.\cite{Zetterling2002}  The ability to engineer the local photonic environment of optically active solid-state qubits through the use of microfabrication techniques is crucial for their scalable application to the field of quantum information.\cite{Loncar2013}  In contrast to the most widely studied forms of SiC (4H and 6H),  the availability of cubic 3C-SiC as a single crystal heteroepitaxial layer on silicon opens up the possibility of combining the favorable properties of SiC defect qubits with the fabrication capabilities available in III-V and Silicon-on-Insulator semiconductor systems. This may enable on-chip, scalable architectures for generating,{\cite{Luxmoore2013}} routing,{\cite{Faraon2011}} manipulating,{\cite{Kennard2013}} and detecting{\cite{Reithmaier2013}} single photon emission from defect qubits in SiC. These capabilities can be combined with  small mode volume optical cavities \cite{Yamada2014} \cite{Radulaski2013} \cite{Lee2015} \cite{Bracher2015} to utilize strong Purcell enhancements{\cite{Faraon2011} \cite{Faraon2012}} for efficient single photon generation or to use cavity QED protocols to realize a spin-photon interface \cite{Li2015} as a node in an optically connected `quantum network'.\cite{Kimble2008} \cite{Pfaff2014}\\  
{\tab}To date, only modest quality factors (`Q') of up to 1,550 have been demonstrated for small mode volume (`$V_m$') cavities ($V_m$ \mytilde $(\lambda/n)^3$) operating at wavelengths relevant for coupling to known defect qubit optical transitions in 3C-SiC.\cite{Calusine2014}  Nevertheless, low- and modest-Q microcavities can be used to improve defect qubit optical readout and control schemes.  More specifically, microcavities can be used to compensate for on-chip power coupling limitations {\cite{Chang2010}} to facilitate optical coherent control schemes,{\cite{Bose2011}} generate on-chip frequency conversion,\cite{McCutcheon2009} and improve rates of absorption and fluorescence.\cite{Shiue2013} \cite{Jensen2014} \cite{Liu2015}  In this work, we utilize cavity resonant excitation to increase the photoluminescence (PL) count rate, optically detected magnetic resonance (ODMR) signal intensity, and rate of spin state initialization for Ky5 defect spin ensembles incorporated into 3C-SiC photonic crystal cavities.  Furthermore, we use these signal improvements and techniques to study inhomogeneity in the Ky5 defect spin and optical properties and extract estimates of the defects' sublevel transition rates.\\  
\begin{figure*}[!ht]
\centerline{\includegraphics[scale=1]{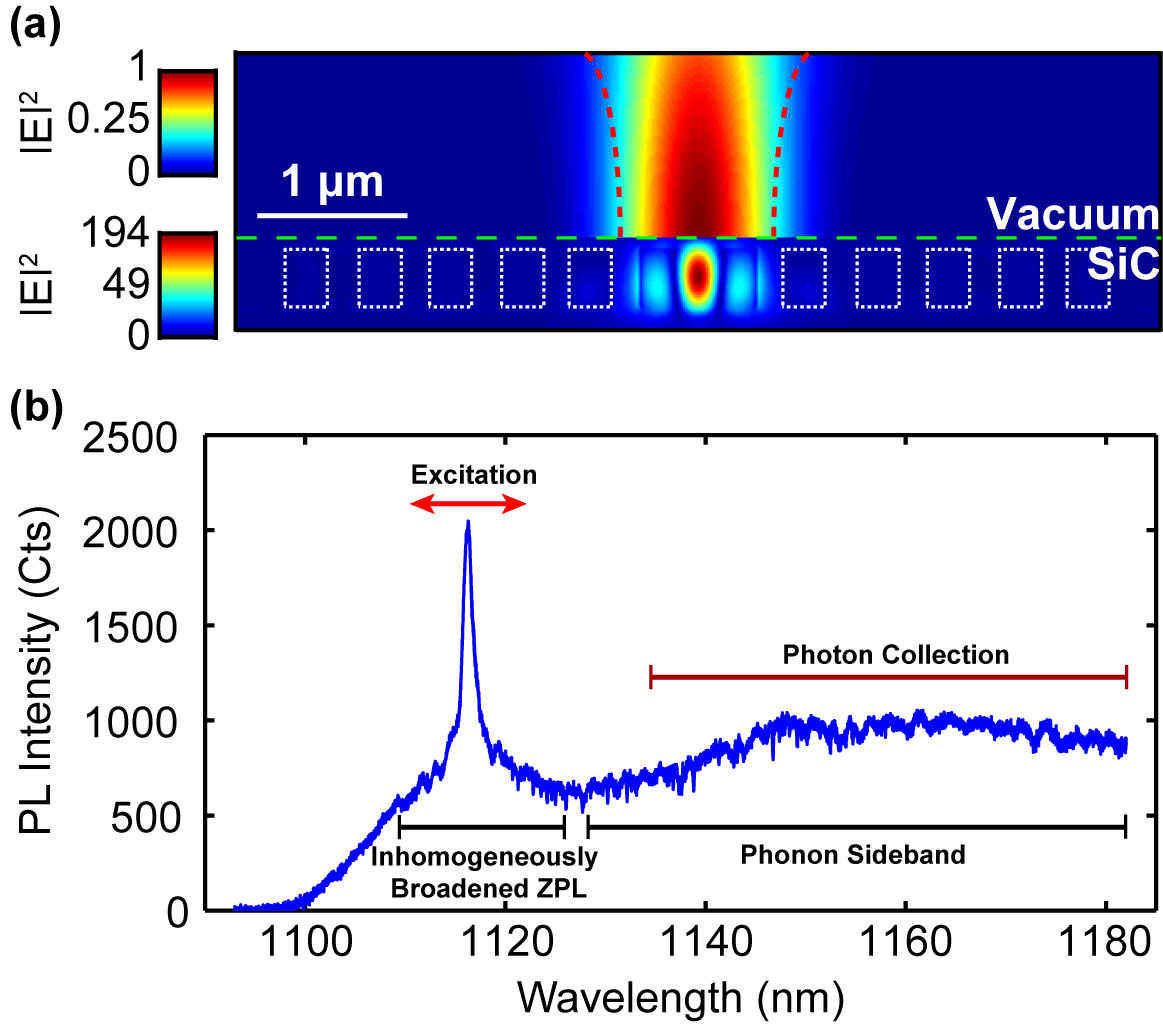}}
\caption{(a) Simulation of the L3 cavity electric field intensity enhancement relative to an incident Gaussian beam (1 $\mu$m beam waist indicated by the dashed red lines) with unity electric field amplitude maximum.  Note the change in scale above and below the dashed green line.  The dashed white lines delineate the cross sections of the photonic crystal cavity holes.  (b) Diagram depicting the CEPLE scheme overlaid on top of the off-resonantly excited PL spectrum of an L3 cavity with incorporated Ky5 centers at 20K.  The wavelength of the fundamental mode of the cavity is near the peak of the inhomogeneously broadened ZPL.}
\label{fig:Fig1}
\end{figure*}
{\tab}The Ky5 point defect is a color center in 3C-SiC that has been previously demonstrated to exhibit spin and optical properties that are similar to the negatively charged nitrogen vacancy (NV) center in diamond. \cite{Falk2013}  It produces an optical emission band consisting of a zero phonon line (ZPL) around 1118 nm and a red-shifted phonon sideband extending out to approximately 1300 nm and, like the NV center, its $S=1$ spin ground state can be initialized and read out optically.  Spin coherence times ($T_2$) in excess of 20 $\mu$s have been demonstrated for the ground-state spin sublevels of Ky5 ensembles.  These states can be manipulated with resonant microwave pulses on fast time scales (tens of nanoseconds) even up to room temperature, making Ky5 centers a viable candidate for a defect-based spin qubit.  The Ky5 center can be controllably generated within the 3C-SiC crystal lattice through a combination of radiation damage and subsequent annealing and has been tentatively identified as a neutral divacancy.\cite{Bratus2009}  Due to the availability of 3C-SiC as a heteroepitaxial film on silicon, photonic structures with three-dimensional optical confinement can be fabricated by removing the underlying substrate through conventional silicon wet and dry etch processes.
\begin{figure*}
\centerline{\includegraphics[scale=1]{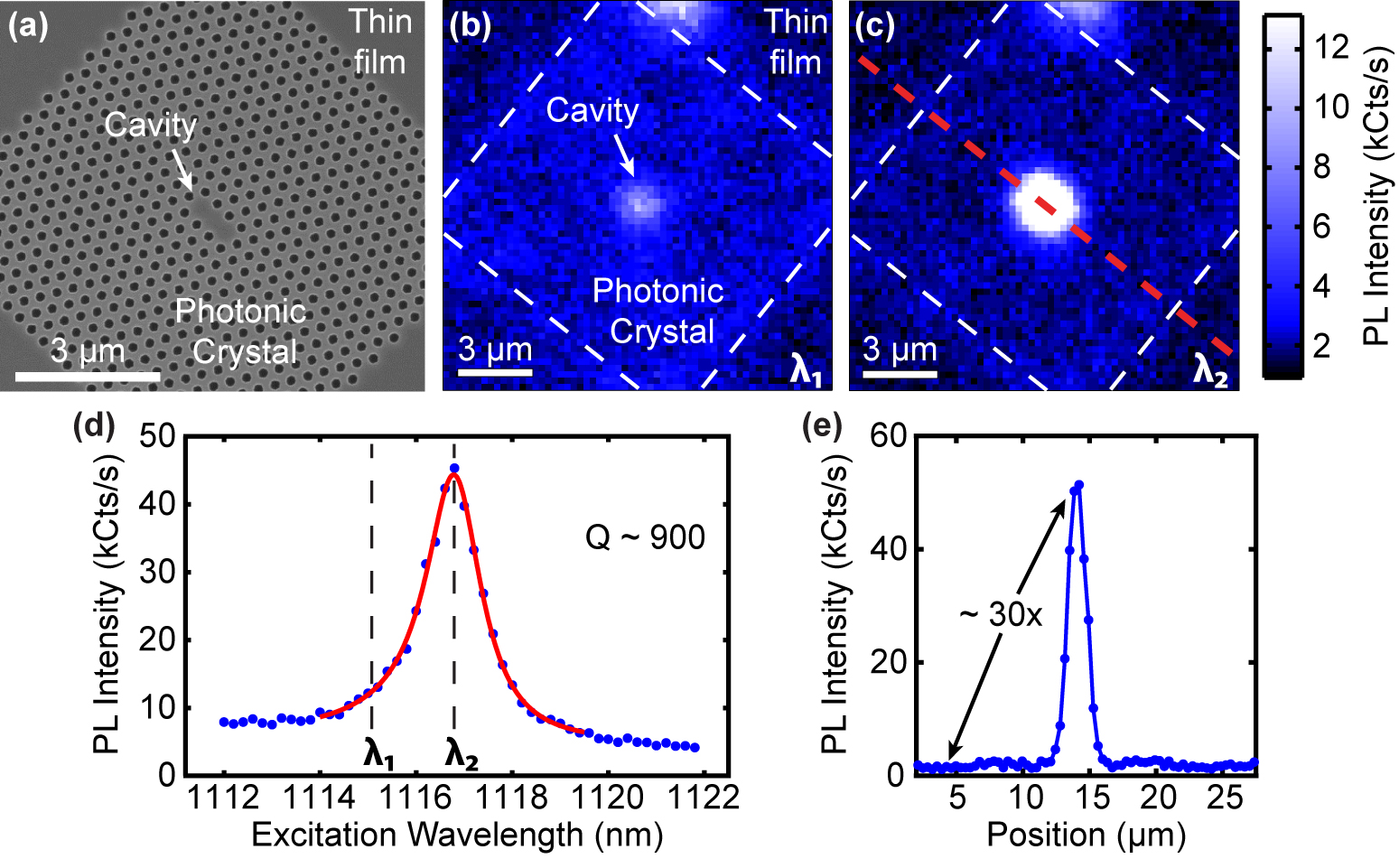}}
\caption{(a) SEM image of the 3C-SiC L3 cavity structure. (b) and (c) 15 $\mu$m by 15 $\mu$m spatial scans of the sideband PL intensity signal for the excitation wavelengths depicted in (d)((b): $\lambda_1$=1115.1 nm, (c): $\lambda_2$=1117.1 nm).  The white dashed lines indicate the extent of the photonic crystal.  The count rate measured on the cavity in (c) was \mytilde 48 kCts/s. (d) Sideband PL count rate vs. excitation wavelength measured at the position of the cavity.  (e) Linecut along the dashed red line  in (c).}
\label{fig:Fig2}
\end{figure*}
\section{\large \RNum{2}. CAVITY-ENHANCED EXCITATION}
{\tab}The use of on-chip photonic cavities to enhance defect qubit excitation is analogous to the use of macroscopic enhancement cavities to generate large intracavity optical fields for applications such as high sensitivity absorption spectroscopy in atomic gases{\cite{Ye2003}} or efficient non-linear optical frequency conversion.  In general, for a fixed incident excitation power, the local electric field intensity inside a microcavity scales as $|E|^2 \propto \frac{\eta Q }{V_m}$  where $\eta$ is the input power coupling efficiency.  More specifically, temporal coupled-mode theory can be used to calculate the response of a photonic crystal cavity subject to far field excitation by an externally incident Gaussian beam, yielding the expression:\\
\begin{equation}
\frac{|E_c|^2}{|E_0|^2}=\frac{Q \eta \lambda n_o w_o^2}{4 n_c^2 V_m}
\end{equation} 
\\
where $|E_c|^2$  is the cavity electric field intensity maximum,  $|E_0|^2$  is the incident beam electric field intensity maximum, $\lambda$ is the wavelength of the cavity mode, $n_o$ is the refractive index of the medium surrounding the cavity, $w_o$ is the incident beam waist, and $n_c$ is the cavity index of refraction (\mytilde2.64 for 3C-SiC).\cite{Supp}  Alternatively, finite-difference time-domain simulations can be used to directly calculate the steady state response of the cavity to continuous wave excitation.  Figure \ref{fig:Fig1}(a) shows the excitation geometry used for simulating the local field enhancement generated in a 300 nm thick L3 cavity for an incident Gaussian electric field intensity with unity amplitude and beam waist equal to our experimentally measured value of \mytilde 1 $\mu$m.  The resulting simulated electric field intensity enhancement of 193.48 agrees within 1\% of the value of calculated from Eq. 1 using the input parameters extracted from simulations for the L3 cavity design.  See \cite{Supp} for a comparison of the local field enhancements and cavity parameters for different structure designs and a more detailed discussion of the simulations. \\
{\tab}To observe these cavity enhanced field intensities, we measured photonic crystal cavities that were designed to exhibit fundamental modes tuned to the inhomogeneously broadened ZPL of the Ky5 centers incorporated into the cavity [around 1118 nm, see Fig. \ref{fig:Fig1}(b)].  The cavities consisted of  L3 and H1 structures fabricated in 300 nm thick 3C-SiC films that were implanted with carbon ions at an energy of 110 KeV and annealed at 750$^o$ C for 30 minutes in order to generate ensembles of Ky5 defects.  Details of the sample design, fabrication, and characterization are presented in Ref. \cite{Calusine2014}.  We primarily focused on L3 designs to observe cavity-enhanced optical excitation due to their greater degree of coupling to far field Gaussian modes as compared to the H1 cavities, which were better suited for improved narrowband collection of off-resonantly excited defect PL (10-20 times improvement over unpatterned thin films).  The measured L3 structures exhibited Q \mytilde 900 with simulated mode volumes of .68 $ (\lambda/n)^3$ and far field coupling efficiencies to external, free space Gaussian modes of 9.6\%.\cite{Supp}\\
{\tab}All measurements were performed in a home-built scanning confocal microscope with an integrated helium flow cryostat at 20K.  A 1064 nm diode laser was used for off-resonant excitation and a tunable (1090-1180 nm), narrow linewidth ({\textless}300 kHz) Littman-Metcalf diode laser was used for resonant photoluminescence excitation and cross-polarized resonant scattering spectroscopy.  Samples were mounted with the cavity axis at 45 degrees with respect to the incident laser polarization to allow for the use of the latter technique for control measurements.  Collected PL and reflected laser light passed back through the objective and were detected using an InGaAs CCD array, a low noise femtowatt photoreceiver, or a superconducting nanowire single photon detector (SNSPD).  See \cite{Supp} for a detailed description of the experimental apparatus and control measurements.\\
\begin{figure*}
\centerline{\includegraphics[scale=1]{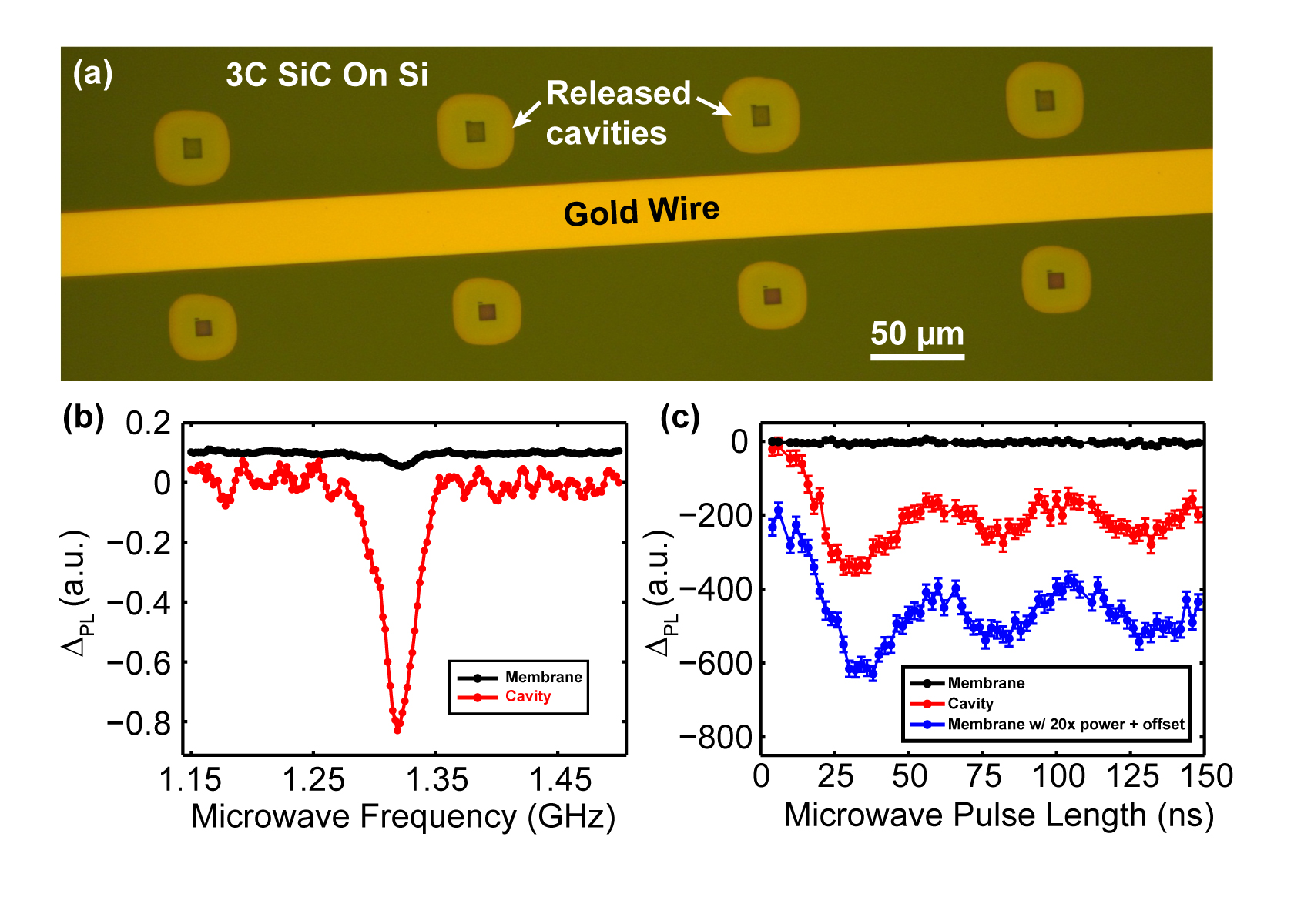}}
\caption{(a) Optical image of a photonic crystal cavity array with a 50 $\mu$m wide 10 nm/300nm Ti/Au microstrip positioned 50 $\mu$m away from the cavities.  (b) ODMR signal at 1.319 GHz measured on the photonic crystal cavity (black line and dots) and on the 3C-SiC thin film (red line and dots) for excitation at the same power at the cavity resonance wavelength. (c) Pulsed ODMR signal from a Ky5 ensemble subject to cavity-enhanced excitation (red line and dots) as compared to an ensemble within the thin film (black line and dots).  The blue line and dots show the same measurement for defects within the thin film for a higher optical excitation power (offset for clarity).}
\label{fig:Fig3}
\end{figure*} 
{\tab} We determined the wavelength of the cavity resonances by measuring the PL spectrum of emitters localized within the cavity under off-resonant 1064 nm excitation as shown in Fig. \ref{fig:Fig1}(b).  Figure \ref{fig:Fig1}(b) also depicts the  excitation and PL collection wavelength ranges of the cavity-enhanced photoluminescence excitation (CEPLE) spectroscopy measurement.  In order to observe enhanced excitation of the Ky5 defects using the cavity mode, we tuned the excitation wavelength to the cavity resonance peak where the light is absorbed by exciting the ZPL transitions and collect red-shifted sideband PL.  To determine the total PL signal count rate increase with respect to the unpatterned, released thin film, we performed a series of spatial scans over the photonic crystal cavity area and compared the overall cavity PL count rate to that of the surrounding thin film.  Figure \ref{fig:Fig2}(a) shows a scanning electron microscope (SEM) image of the L3 cavity and Fig. \ref{fig:Fig2} (b) and (c) show a pair of scanning confocal PL images of a 12 $\mu$m by 12 $\mu$m L3 photonic crystal cavity with excitation wavelengths as designated in the Fig. \ref{fig:Fig2}(d).  The photonic crystal extent is delineated by the white dashed lines and the bright spot in its center is fluorescence originating from the cavity.  Figure \ref{fig:Fig2}(d) shows the excitation wavelength dependence of the PL count rate originating from the cavity location.  The excitation wavelength-dependent count rate matched the cavity mode spectrum and the peak exhibited an approximately factor of 5 increase over off-resonant excitation.  Figure \ref{fig:Fig2}(e) shows a line cut of the PL map corresponding to the red dashed line in Fig. \ref{fig:Fig2}(c).\\
{\tab} At the excitation wavelength corresponding to the cavity resonance, the PL count rate was approximately $\Gamma \approx$ 30 times higher than the PL count rate from the unpatterned thin film, where we have defined $\Gamma$ as the ratio of the resonantly excited cavity PL count rate to the thin film PL count rate at the same excitation wavelength and power.  For H1 designs, we observed a lower maximum $\Gamma \approx $ 13 due to a smaller degree of input coupling for an incident Gaussian beam.  Scattered laser light added a negligible contribution to the measured signal (see Ref. \cite{Supp}).  While we excited cavity modes tuned to the Ky5 defect ZPL wavelength range, this same approach can be applied to applications that require efficient off-resonant excitation\cite{Clevenson2015} and would be particularly beneficial for excitation wavelengths that overlap weakly with the defects' absorption spectrum.\\
{\tab} Aside from increasing the overall luminescence count rate, cavity resonant excitation can also be used to improve measurements on single emitters or ensembles by exciting a significantly smaller sample volume than a standard objective lens configuration.  The excitation volume of a thin film is greatly reduced as compared to that of bulk material because the excitation volume dimension along the optical axis is set by the thickness of the thin film rather than the diffraction limit (\mytilde 10 $\mu$m for a .7 NA objective at $\lambda$=1.1 $\mu$m).  Furthermore, for our optical configuration, the cavity provides a further 12.4-fold reduction in the sample excitation volume as compared to the thin film, resulting in an overall 705-fold reduction as compared to bulk material.\cite{Supp}  This reduction in the excitation of localized states in the  proximity of isolated single emitters within the surrounding material can improve the performance of single photon sources by reducing background fluorescence\cite{Nomura2006(4)} or charge-induced spectral diffusion.\cite{Jelezko2002} 
\section{\large \RNum{3}. OPTICALLY DETECTED MAGNETIC RESONANCE SIGNAL ENHANCEMENTS}
{\tab} The observed signal improvements provide a means to greatly increase the ODMR signal amplitude in order to probe the Ky5 center's spin-dependent electronic structure or for sub-diffraction limit, on-chip sensing applications.  In order to perform spin-dependent measurements on defects within the cavity structure, we performed an additional fabrication step that adds a 10 nm/300 nm Ti/Au metallization layer to the sample surface for applying intense local microwave fields to the sample.  Figure \ref{fig:Fig3} (a) shows an optical image of an array of released 3C-SiC films patterned with photonic crystal cavities next to a 50 $\mu$m wide microstrip positioned 50 $\mu$m away from the structures.  Due to the robustness of the approximately 40 $\mu$m by 40 $\mu$m freestanding films, the metallization can be applied prior to or after the membrane release step without the need for critical point drying. We used on-chip microstrips in order to apply sufficiently intense microwave fields to achieve coherent spin manipulation (Rabi oscillations) on time scales faster than the Ky5 defects' $T_2^*$ of {\mytilde} 50 ns.{\cite{Falk2013}}\\
{\tab} Figure \ref{fig:Fig3}(b) compares the Ky5 ODMR signal ($\Delta_{PL}$) with the excitation beam incident on the released 3C-SiC thin film (black line and dots) and the photonic crystal cavity (red line and dots) for the same optical power at zero magnetic field.  The overall ODMR signal increase matches the PL count rate increase observed for this cavity ($ \Gamma \approx$ 20), confirming that the PL count rate increases under resonant excitation originate from defect PL rather than scattered excitation.  Similar improvements were observed for pulsed ODMR signals, as depicted in Fig. \ref{fig:Fig3}(c).  By synchronizing pulsed optical excitation to polarize and readout the spin ensemble with pulsed microwaves for spin manipulation, we observed Rabi oscillations between the $m_s=0$ and $m_s= \pm1$ spin sublevels of the defects' ground states for emitters excited via the cavity mode or in the unpatterned thin film.  For pulsed ODMR measurements, signal is greatly increased under resonant excitation due to a combination of both higher PL count rates and faster optical pumping.\\
{\tab}CEPLE spectroscopy on the inhomogeneously broadened ZPL of ensembles can greatly increase measurement signal intensities and reduce the number of measured emitters to facilitate studies of the optical and spin inhomogeneity that we observe for defects in 3C-SiC thin films.\cite{vanOort1992} \cite{Calusine2014} \cite{Supp} For high-emitter-density samples ($10^{13}$ ions cm$^{-2}$ implantation dose)  like those measured in the Fig. \ref{fig:Fig3}, we observed variations in the defect ground-state zero field splitting (D $\approx$ 1.32 GHz) of up to 10 MHz and variations in the ensemble ODMR linewidth by almost a factor of 2 as the excitation wavelength is varied within the inhomogeneously broadened ensemble ZPL (FWHM of 28.2 nm).  For lower-density samples ($10^{12}$ ions cm$^{-2}$ dose), the 10-20 times improvement of collection efficiency \cite{Calusine2014} for defects emitting into the cavity mode reveals PL lines as narrow as 25 GHz within the inhomogeneously broadened ZPL (28.2 nm FWHM $\approx$ 6.7 THz).  These emission lines were also observed with CEPLE spectroscopy for significantly lower excitation powers (1.5 $\mu$W vs. 3.2 mW) and higher PL count rates (\mytilde 1,400 cts/s vs. \mytilde 200 cts/s) as compared to narrowband PL collection of the emission line using a scanning monochromator.\cite{Supp}  Exciting and collecting PL from a reduced mode volume facilitates the isolation of a smaller number of spectral lines within a given wavelength range because this spectral fine structure and its polarization can vary throughout the thin film, likely due to variations in the local crystalline environment. \cite{Bosi2013} 
\section{\large \RNum{4}. ENHANCED SPIN INITIALIZATION RATES}
\begin{figure*}
\centerline{\includegraphics[scale=1]{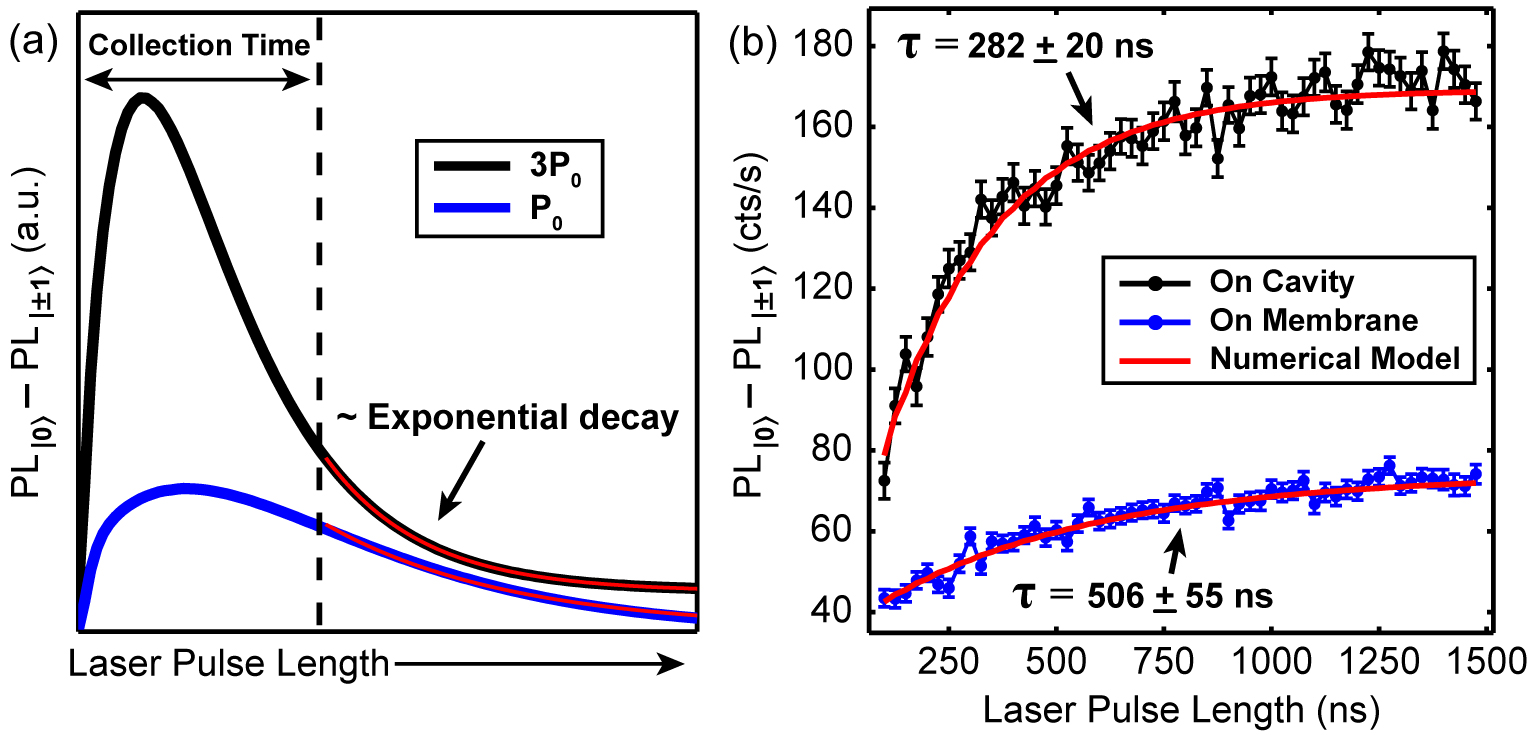}}
\caption{(a) Simulated time-dependent difference in PL for an NV center initialized into the $m_s=0$ vs. $m_s=\pm 1$ ground state for two excitation rates differing by a factor of 3.  The red lines depict exponential fits to the time scales that correspond to an approximately exponential saturation of the ground-state spin polarization. The black curve is offset for clarity.  (b)  Measured PL difference signal for an ensemble of Ky5 defects within an L3 photonic crystal cavity excited on resonance (black line and dots) as compared to an ensemble excited within the unpatterned thin film (blue line and dots) at the same power and wavelength.  The red lines represent simulated behavior as predicted by the model discussed in the main text.  The time constants result from exponential fits to the data. }
\label{fig:Fig4}
\end{figure*} 
{\tab}In addition to providing increased signal intensities, the excitation enhancement provided by the resonant cavity mode can also be used to increase the rate of optically-induced spin polarization in the Ky5 defects' ground state.  To observe this process, we measured the defects' time-dependent PL intensity in response to variable-length laser pulses with interleaved microwave $\pi$ pulses driving the $m_s=0$ to $m_s= \pm 1$ ground-state spin sublevel transitions.  A series of optical pulses with a variable length from 250 to 1500 ns were applied to excite the Ky5 defects from the ground state to the excited state where they experience either spin-conserving radiative recombination or spin-dependent relaxation to the ground-state spin sublevels predominantly via the intersystem crossing (ISC) transitions.\cite{Manson2006}  As a result of this relaxation pathway, optical excitation achieves ground-state spin polarization at a rate that depends on the optical excitation pulse length and field intensity.  Relaxation through the ISC transitions also provides a means for reading out the spin state: for a period of time immediately after the turn-on of the optical pulse, the Ky5 defects will fluorescence more or less brightly depending on the ground-state spin polarization immediately prior to the optical pulse.  Accordingly, we collected the PL for a fixed time period of 100 ns immediately after the turn-on of the optical pulse to obtain a measure of the spin polarization generated by the previous pulse.  Prior to half of the optical pulses, we applied a microwave $\pi$ pulse to invert the ground-state spin polarization, which was then re-pumped to the steady state polarization by the subsequent optical excitation.  We measured this difference in PL (PL$_{\mid 0 \rangle} - $PL$_{\mid \pm{1} \rangle}$) between optical pulses that immediately followed a $\pi$ pulse and those that did not to obtain a measure of the ground-state spin polarization.  The optical excitation was turned off for at least 500 ns prior to microwave manipulation and readout to allow for population within the ISC levels to fully relax to the ground state.  See \cite{Supp} for pulse sequences and further details.\\
{\tab} The same model used to explain the PL dynamics of the NV center in diamond can be applied to model the expected dynamics.\cite{Manson2006}  Figure \ref{fig:Fig4}(a) shows the results of numerical simulations of the time-dependent difference in PL between an NV center that has been initially prepared in the $m_s=0$ state and the $m_s= \pm 1$ states for two excitation rates differing by a factor of 3.  The figure also depicts the measurement scheme described above and the approximately exponential decay of the PL difference curves for laser pulse lengths longer than the collection time (red lines).  This behavior corresponds to an approximately exponentially saturating ground-state spin polarization and PL difference signal.  We observed corresponding behavior when comparing the laser pulse length-dependent PL difference signal for Ky5 defects excited within the thin film with those excited using the photonic crystal cavity mode, as shown in Fig. \ref{fig:Fig4}(b).  Figure \ref{fig:Fig4}(b) also shows the results of numerical simulations of this difference signal for varying laser pulse lengths.  The observed saturation behavior matches the predicted dynamics for intersystem crossing rates that differ from the NV center, as determined from a nonlinear least squares fit to a numerical model that includes the expected excitation enhancement ($|E_c|^2$/$|E_0|^2$ \mytilde 96) for this structure geometry.\cite{Supp}  The enhanced excitation rate for defects within the cavity structure generates a faster rate of ground-state spin initialization corresponding to a shorter time constant $\tau$.  The time constant associated with this approximately exponential saturation of the PL difference signal decreases from 500 $\pm$ 55 ns to 280 $\pm$ 20 ns.  This approximately factor of 2 difference in the time constant for optical pumping is relatively small as compared to the overall PL count rate increase (\mytilde 20x) because for high excitation rates, the spin initialization rate becomes limited by the transition rate to the intersystem crossing levels rather than the rate of optical cycling.  As a control measurement to verify the spin-dependent origin of these dynamics, we removed the microwave $\pi$ pulse and observed no difference signal. \cite{Supp} \\
{\tab}These measurements provide the first direct observation of spin-dependent PL dynamics in SiC and support previous assumptions that these defects exhibit optical dynamics similar to the NV center in diamond.  Furthermore, we corroborated the variable length laser pulse results using TCSPC methods and observed a spin-dependent pulsed ODMR contrast of 2.75\%.  By combining this value with the numerical modeling, we estimate an intrinsic spin-dependent pulsed ODMR contrast of 4.9\% for ideal measurement parameters (for details, see Ref. \cite{Supp}).  While this value is lower than typically observed for single NV centers, it is similar to what has been previously observed for high density NV center ensembles and single defects in SiC. \cite{Maertz2010} \cite{Christle2014}
\section{\large \RNum{5}. CONCLUSIONS AND OUTLOOK}
{\tab}In conclusion, we have demonstrated resonant excitation of 3C-SiC photonic crystal cavities with integrated defect spins for large PL and ODMR signal enhancements and increased spin initialization rates.  Our analysis shows that our present cavity designs are capable of achieving localized optical field intensities that can be enhanced by a factor of almost 200 relative to an incident Gaussian beam.  This value could be further improved with concurrent optimization of cavity Q and coupling to the Gaussian excitation mode.\cite{Portalupi2010}  For measurements of ensembles of defects, these small mode volumes and excitation intensity enhancements may facilitate on-chip applications that are limited by inefficient optical coupling or poor spectral overlap of the excitation source and defect absorption bands.  These applications include spin ensemble-based sensing techniques using on-chip optical traps,\cite{VanLeest2013} enhanced absorption for hole-burning experiments,\cite{Harley1984} or for enhancing the signal of spectrally distinct sub-ensembles in order to study inhomogeneous broadening.\cite{vanOort1992}  For applications involving single defects, cavity resonant excitation can provide enhanced optical stark shifts,\cite{Bose2011} compact, on-chip single photon frequency conversion,\cite{McCutcheon2009} or reduced excitation of background impurities that are detrimental to single photon source performance.\cite{Nomura2006(4)} \cite{Bassett2011}  Additionally, we have provided evidence that defects in 3C-SiC exhibit photodynamics similar to those of the NV center in diamond.  These results underscore the benefits of fabricating devices with integrated defect spin states in heteroepitaxial 3C-SiC as a means to incorporate defect qubits into scalable device architectures for applications in the field of quantum information and sensing.\\
\\
{\tab}We thank David Christle, Bob Buckley, and Joerg Bochmann for helpful discussions.  This work was supported by the AFOSR QuMPASS MURI FA9550-12-1-004 and NSF DMR-1306300.  A portion of this work was done in the UC Santa Barbara nanofabrication facility, part of the NSF funded NNIN network.  We acknowledge support from the Center for Scientific Computing from the CNSI, MRL: an NSF MRSEC (DMR-1121053) and NSF CNS-0960316.

\end{document}